\documentclass[aps,pra,twocolumn,superscriptaddress,showpacs,longbibliography,10p]{revtex4-1}

\usepackage[utf8]{inputenc}

\usepackage{graphicx}
\graphicspath{{./fig/}{./}{./fig_supp/}}
\usepackage[hidelinks]{hyperref}
\usepackage{amsmath}
\usepackage{amssymb}
\usepackage{epstopdf}
\usepackage{color}
\usepackage[normalem]{ulem}
\usepackage{changepage}

\newcommand{\red}[1]{{\color{red}#1}}

\newcommand{\ket}[1]{{\vert#1\rangle}} %ket
\newcommand{\abs}[1]{\left|#1\right|} %abs
\newcommand{\mean}[2] {  \langle  #1 \rangle _{#2} }

\newcommand{\Rb}{$^{\text{87}}\text{Rb }$}

\newcommand{\F}[1]{\left| F=#1 \right\rangle}
\newcommand{\FPrime}[1]{\left| F'=#1 \right\rangle}
\newcommand{\Fmf}[2]{\left| F=#1, m_F=#2 \right\rangle}

\newcommand{\citel}[1][]{%
	\red{\ifthenelse{\equal{#1}{}}{[?]}{[#1]}}%
}

\renewcommand{\Im}{\operatorname{Im}}
\renewcommand{\Re}{\operatorname{Re}}

\renewcommand{\part}[1]{}

\newcommand{\Fig}{Fig.\@ }
\newcommand{\Eq}{Eq.\@ }
\newcommand{\Ref}{Ref.\@ }

\newcommand{\ShowTitle}[1]{#1}

\usepackage{array}
\newcolumntype{L}{>$l<$}
\usepackage{units}
\usepackage{setspace}
\usepackage{xspace}
\usepackage{tikz}

\renewcommand{\part}[1]{\paragraph{\textbf{#1}}}

\begin{document}

\title{Deterministic Shaping and Reshaping of Single-Photon Temporal Wave Functions}

\author{O. Morin,  M. Körber, S. Langenfeld, and G. Rempe}
\affiliation{Max-Planck-Institut f\"{u}r Quantenoptik, Hans-Kopfermann-Strasse 1, 85748 Garching, Germany}
\date{\today}

\begin{abstract}
Thorough control of the optical mode of a single photon is essential for quantum information applications. We present a comprehensive experimental and theoretical study of a light-matter interface based on cavity quantum electrodynamics. We identify key parameters like the phases of the involved light fields and demonstrate absolute, flexible, and accurate control of the time-dependent complex-valued wave function of a single photon over several orders of magnitude. This capability will be an important tool for the development of distributed quantum systems with multiple components that interact via photons.
\end{abstract}


\maketitle

%\part{Introduction}

Single photons are of paramount importance for modern quantum information science. Envisioned applications range from all-optical quantum computation \cite{Knill2001,Minzioni2019} to quantum communication in nonlocal quantum clouds \cite{Kimble2008,Wehner2018}. As all the conceived quantum information protocols involve, in one way or another, interference effects, coherent photons are mandatory for the implementation of these protocols \cite{Rohde2005,Raymer}. This requirement includes the (relative) coherence within the photon wave packet as well as the (absolute) coherence with respect to a common network reference clock. Full control over amplitude and phase of the photon's temporal mode is a challenge \cite{Keller2004,Eisaman2004,Eisaman2005,Balic2005,Kolchin,Kielpinski2011,McKinstrie2012,Matsuda2016,Fisher2016,Karpinski2017,
Zhou2012,Specht2009,NisbetJones2011,NisbetJones2013,Farrera2016,Averchenko2017,Sych2017}, as is the mode conversion for all regimes from narrow to broad-bandwidth single photons. Meeting these challenges would open up new possibilities like temporal mode matching in multimode quantum networks or information encoding in the optical mode of the photon.

Here we exploit the arsenal of cavity quantum electrodynamics (CQED) \cite{Reiserer2015} and demonstrate deterministic mode control over single optical photons. Toward this end, we extend previous models \cite{Law1997,Kuhn1999,Fleischhauer2000, Gorshkov2007,Muecke2013,Dilley2012,Dalibard92,Giannelli2018} and take into account the full energy-level structure of the atom that serves as photon emitter and photon receiver in a high-finesse optical resonator. We find a surprisingly strong frequency dependence of the process efficiency with a pronounced minimum that originates from destructive interference of transition amplitudes. We also show that the emission efficiency is not a reliable measure for photon coherence. We moreover shape the photon phase, demonstrate the mode selectivity of photon absorption, and stretch and compress a given single-photon wave packet by 3 orders of magnitude. All these achievements are realized in combination with a convenient-to-implement coherence-testing method that outputs the time-dependent complex-valued temporal mode function with minimal resources \cite{Morin2019}. 
As both the amplitude and the phase of the temporal mode are determined with respect to a commonly accepted reference, a phase-locked laser, we are now in a position to certify the (absolute) coherence of a network photon.

\begin{figure}[!b]
\includegraphics[width=0.99\columnwidth]{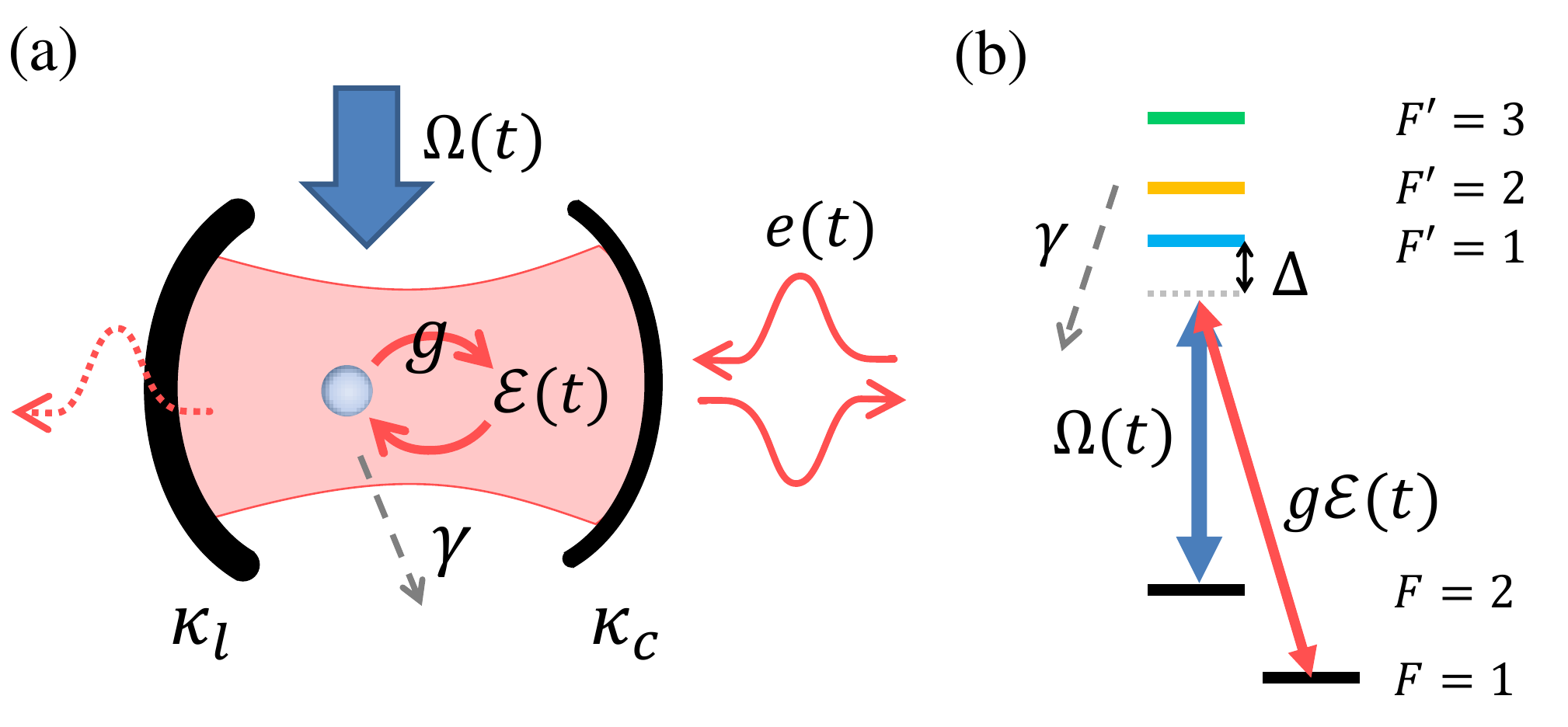}
\caption{Model of the experimental setup. (a) A single atom of $^{87}$Rb is trapped in a high finesse optical cavity via a 2D optical lattice (not shown). The control field of Rabi frequency $\Omega(t)$ and $\pi$-polarization impinges from the side of the cavity. The incoming photon is a weak coherent pulse of temporal shape $e(t)$. The atom is coupled to the cavity field $\mathcal{E}(t)$ with the light-matter coupling constant $g$. (b) We use a $\Lambda$ scheme with three excited-states. The Zeeman states involved are $m_F=0$ for $F=1$ and $m_F=-1$ for $F=2$ and $F'=1,2,3$.}
\label{fig:setup}
\end{figure}

%\part{Setup}
Our system uses the quantum memory scheme described in \Ref \cite{Koerber2018} and therefore all the following results are also valid for single photons encoding a qubit in their polarization degree of freedom. We use a single $^{87}$Rb atom trapped in an optical high-finesse cavity. The cavity is resonant with the $\text{D}_2$ line of $^{87}$Rb at 780nm. Our output (or input) photon is resonant with the cavity and addresses the transition $\ket{F=1,m_F=0} \leftrightarrow \ket{F'=1,m_F=-1}$ with a single-photon detuning $\Delta$ (see \Fig \ref{fig:setup}). The control light addresses the transition $\ket{F=2,m_F =-1} \leftrightarrow \ket{F'=1,m_F =-1}$  with the same detuning $\Delta$ such that the $\Lambda$ scheme is in two-photon resonance. %\blue{Ideally the two-photon detuning is $\delta=0$}. 
The combination of the two light fields drives the atomic population from the state $\ket{F=1, m_F=0}$ to $\ket{F=2,m_F=-1}$ in absorption and vice versa for emission. The cavity is asymmetric such that the photon mostly enters/leaves the cavity via the mirror with the highest transmission. We refer to the decay rate via this mirror as $\kappa_c$, whereas we name $\kappa_l$ the decay rate via the second mirror including intracavity optical losses  (see \Fig\ref{fig:setup}).

%\part{Theory}
\Ref \cite{Giannelli2018} provides a comparison of different models describing the photon absorption/emission by atomic systems \cite{Fleischhauer2000,Dilley2012,Gorshkov2007,Dalibard92}. The one proposed by Gorshkov \textit{et al.} \cite{Gorshkov2007} is particularly relevant for us as it considers a single-photon detuning $\Delta$. We adapt the derivation to a system with multiple excited states and obtain two important expressions, one for the relation between the complex-valued temporal shape of the single photon and the control field and another one for the efficiency of the absorption/emission.

The shape of the single-photon state is specified by its  temporal mode function $e(t)$ i.e. $\ket{1}=\int_\mathbb{R}dt\ e(t) \hat{a}^\dagger(t)\ket{0}$.
For the emission process, the link between the single-photon shape $e(t)$ and the Rabi frequency of the control field $\Omega(t)$ is of the same form as the one derived in \cite{Gorshkov2007}: 
%with respect to the time variable 
\begin{multline}\label{eq:omega}
\Omega(t) = \frac{e(t)}{ \sqrt{ 2 \Re[K] \int_t^\infty |e(t')|^2 dt' }} \\
	\exp\left(- i \frac{\Im[K]}{2\Re[K]} \ln \left( \int_t^\infty |e(t')|^2 dt' \right) \right). 
\end{multline}
Here $K$ is a parameter containing all the characteristics of the atomic structure, the cavity parameters and the detunings, as derived in the Supplemental Material (SM). For coherent light-matter interaction, time reversal guarantees that the emission and the absorption processes are described by the same equation: the efficiency is the same and \Eq (\ref{eq:omega}) written for emission is just reversed in time and complex conjugated for the  absorption.

\begin{figure}[!b]
\includegraphics[width=0.99\columnwidth]{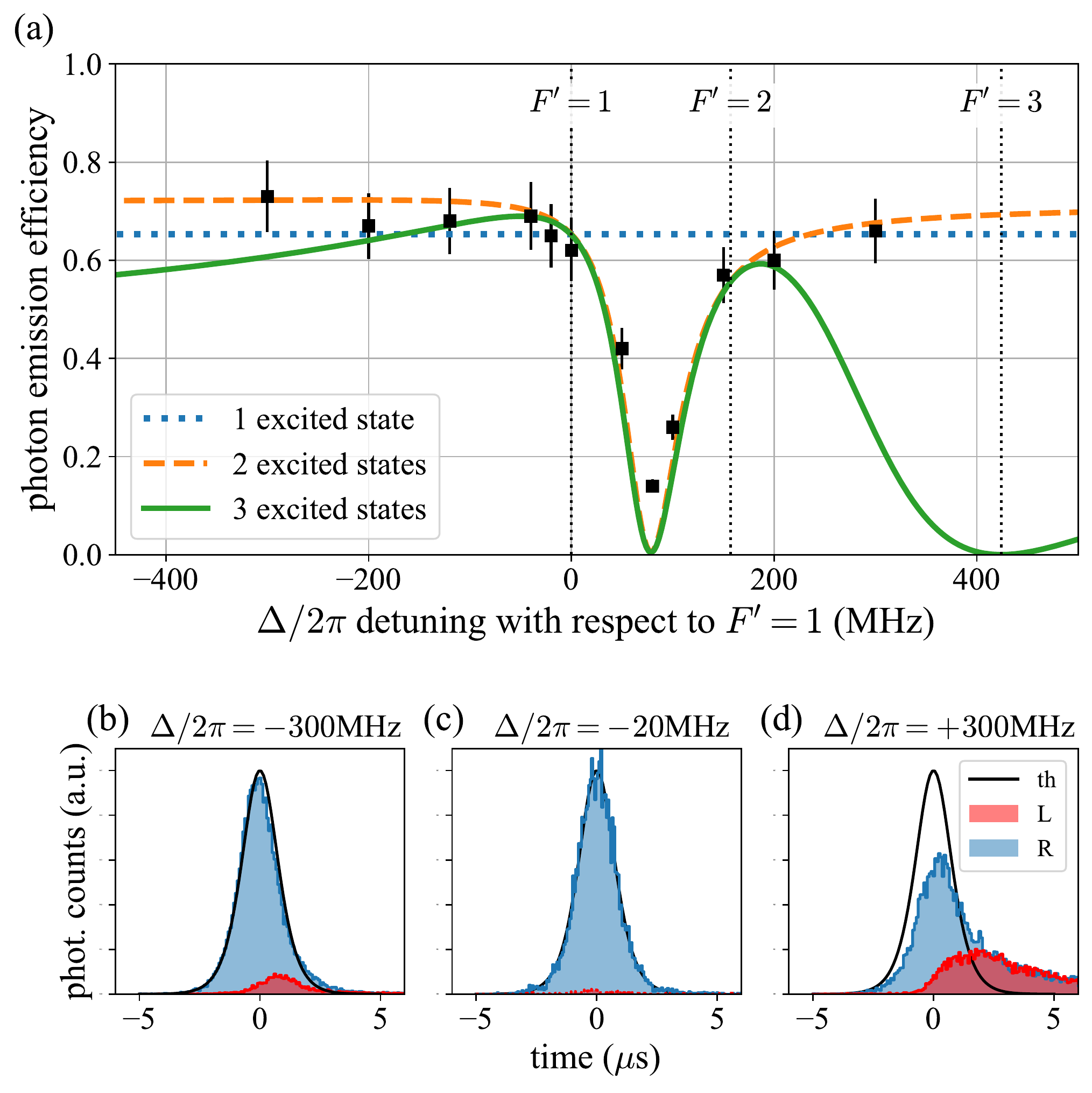}
\caption{Efficiency as a function of the single-photon detuning. (a) The three curves correspond to three theoretical models: The dotted blue line for one excited state with $F'=1$, the orange dashed line for two excited states with $F'=1,2$ and the plain green line for three excited states. The dots correspond to experimental measurements. The statistical errors are smaller. Data can be reproduced to within about 10\% (error bars). (b)-(d) Normalized temporal amplitudes for different detunings. The black lines correspond to the targeted shapes. The blue histogram corresponds to the detected photons with right circular polarization (R) and the red with left circular polarization (L). The amount of left polarized light compared to the expectation of only right polarized light witnesses the increase of incoherent processes for large detunings.}
\label{fig:efficiency}
\end{figure}

Following \cite{Gorshkov2007}, the ideal efficiency of the photon absorption/emission process $\eta_C=2C/(2C+1)$ only depends on the cooperativity $C=g^2/2\kappa\gamma$, where $g$ is the light-matter coupling constant, $\kappa = \kappa_c+\kappa_l$ the cavity field decay rate and $\gamma$ the atomic polarization decay rate. For our setup, the parameters have the values $(g,\kappa_c,\kappa_l,\gamma)=2\pi\times(4.9,2.4,0.3,3.03)\ $MHz \cite{CQEDNote}. For an imperfect cavity, however, the overall efficiency reduces to the product of $\eta_C$ and the escape efficiency $\eta_\text{esc}=\kappa_c/(\kappa_c+\kappa_l)$ \cite{Muecke2013}. With multiple excited states, the efficiency becomes a function of the single-photon detuning, $\eta_C(\Delta)$. Indeed, $\Delta$ modifies the strength of the interaction with each excited state.

%\part{Efficiency}
Figure \ref{fig:efficiency} (a) compares the experimentally measured efficiency with three different models. The three curves correspond to the models with $F'=1$ only, with both $F'=1$ and $F'=2$, and all three excited levels $F'=1,2,3$. 
Both states $\ket{F'=1}$ and $\ket{F'=2}$ participate in the Raman process. Depending on the detuning, constructive or destructive interference occurs in the photon absorption/emission process. This explains that the efficiency varies with the detuning and can be larger or smaller than in the case with only one excited level $F'=1$. In contrast, $\ket{F'=3}$ does not couple to the ground state $\ket{F=1}$ but potentially destroys the emission process by incoherent scattering.

Surprisingly, the experimental data tend to follow the model with two excited levels. This is explained by the fact that the models provide the efficiency for a pure temporal mode whereas the single-photon counting modules detect any mode. Therefore, the theoretical efficiency can only be compared with the experimental one in the regime of coherent emission.
For instance, populating state $\ket{F'=3}$ in the emission process leads to decay back into the initial $\ket{F=2}$ ground state and thus to a new emission attempt. This starts at a random time and potentially from a different $m_F$ state. On average, this results in a mixture of wavepackets which are all detected by the single-photon counting modules although they are not part of the coherent emission process.

The presence of incoherent processes in the photon emission is well illustrated in \Fig \ref{fig:efficiency} (b)-(d). Here the temporal distributions of the single-photon detection events are plotted for left and right circular polarizations. For a large red detuning or close to the transition to $\ket{F'=3}$, the photon shape alters and (unwanted) left circularly polarized photons are generated. A simulation of the full system has been performed and supports the experimental observations (see SM).

%\part{Shape emission}

\begin{figure}[!t]
\includegraphics[width=0.99\columnwidth]{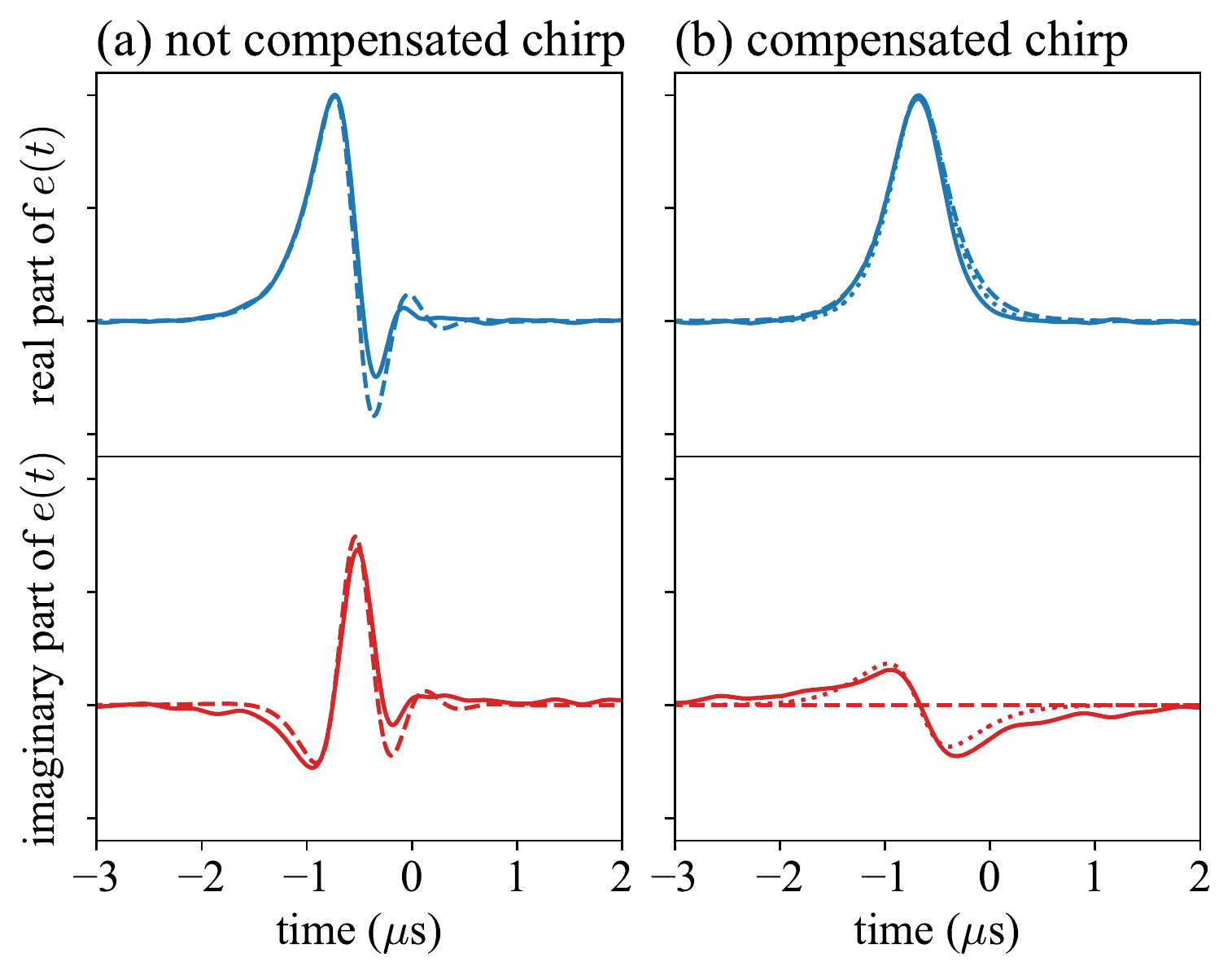}
\caption{Measurement of the complex temporal shape. The real and imaginary parts of $e(t)$ reconstructed by temporal mode analysis are plotted in arbitrary units but identical scales. (a) corresponds to the case without compensation of the phase chirp induced by light shift, and (b) to the case with phase compensation.  The plain lines correspond to the experimental measurements and the dashed lines to the theoretical shapes. When the phase chirp is compensated, the fidelity between the target shape and the measured one is 90\%. The dotted lines in plot (b) correspond to the shape with a residual detuning of 180kHz with which the fidelity goes up to 98\%. In contrast, without phase compensation, the fidelity is 55\%. For these measurements, the detuning is $\Delta/2\pi = -20$MHz and the targeted temporal shape $e(t)$ of the single photon is a hyperbolic secant with the characteristic time $T=0.5\mu$s.
}
\label{fig:temporal_shape}
\end{figure}

The presence of light in both polarization modes indicates the occurrence of incoherent processes. \textit{A contrario}, the absence of two polarizations in \Fig \ref{fig:efficiency} (c) does not guarantee the coherence of the wavepacket. In addition, the measurement via photon counting does not provide any information about the phase of the temporal shape $e(t)$. In particular, \Eq (\ref{eq:omega}) predicts a time-dependent phase term. This term actually accounts for the light shift induced by the control field. Without compensation of this phase, a frequency chirp is imprinted on the emitted photon.
This chirp is unwanted and can enlarge the photon bandwidth to a value larger than the cavity bandwidth, thus decreasing the emission efficiency. We emphasize here that \Fig \ref{fig:efficiency} was measured with phase compensation that is achieved by properly controlling the phase of the control laser.

In order to evaluate the temporal shape $e(t)$ of the emitted photon, i.e. amplitude and phase, we use a temporal mode analysis technique \cite{MorinMode,Morin2019,LvovskyMode}. As explained in the SM, from a set of homodyne measurements we obtain the real and imaginary part of the temporal mode function $e(t)$. Figure \ref{fig:temporal_shape} shows two cases: (a) without phase control and (b) with phase compensation. Figure \ref{fig:temporal_shape}(a) clearly shows the frequency chirp induced by the control field.
The fidelity with a purely real temporal shape is 55\% in that case \cite{Fid_temp_mode}, showing the detrimental impact of the phase chirp in case it is not properly compensated. With compensation, the imaginary part in \Fig \ref{fig:temporal_shape}(b) becomes  smaller and the fidelity increases up to 90\%. The majority of the remaining infidelity is likely to originate from a residual constant detuning of 180kHz of the local oscillator, as including this into the theory results in a fidelity of 98\%.

In addition, we obtain a photon-number distribution with $p_\ket{0}=0.716(2)$, $p_\ket{1}=0.284(2)$ and $p_\ket{2}=0.001(2)$. Considering the global detection efficiency of 0.6 and the atom preparation efficiency of 0.74, we estimate $p_\ket{1}\approx 0.64$ at the output of the cavity for each successful preparation of the atom. This result agrees with the one obtain by single-photon counting (see SM).

\begin{figure}[!t]
\includegraphics[width=0.95\columnwidth]{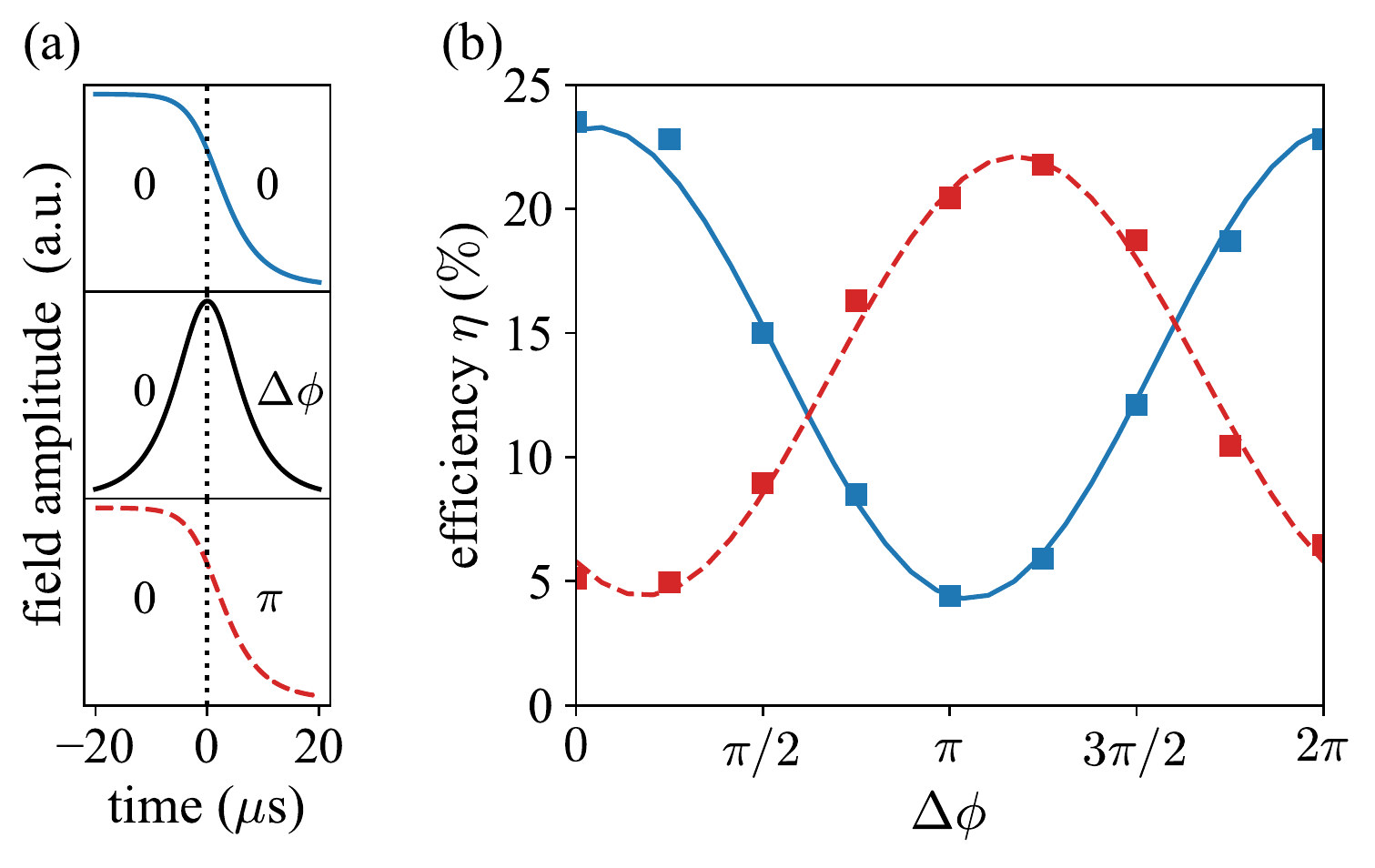}
\caption{(color online) Temporal mode selection. (a) In black is plotted the amplitude of the input photon (experimentally a weak coherent state), which at time $t=0$ has a phase jump of $\Delta\phi$. The two other plots correspond to the amplitude of the control field, in plain blue without a phase jump, in red dashed with a phase jump of $\pi$. (b) The global efficiency (absorption and emission) as a function of $\Delta\phi$ for the two different control phase profiles. The curves are fits of the form $A\sin^2 (\Delta\phi+\phi_0) +B$. }
\label{fig:selection}
\end{figure}

%\part{Mode selectivity}
Similarly to the emission case, the phase of the temporal profile is also important in absorption. Indeed, not properly compensating the phase chirp induced by the light shift leads to a temporal mode mismatch between the incoming photon and the storage mode defined by the shape of the control field. Figure \ref{fig:selection} illustrates this important aspect. We apply a phase jump of $\Delta\phi$ in the middle of the temporal shape of the input photon. Hence, for a $\pi$ phase jump, the input single photon has a temporal shape orthogonal to the one defined by the control field. This results in a suppressed efficiency. In the case depicted in dashed red, we apply a $\pi$ phase jump on the control field such that, this time, the input photon without phase jump should not be stored as it lives in an orthogonal temporal mode. Note that here and in the following experiments, we use a weak coherent state with a well controlled shape as an input.

\begin{figure}[!t]
\includegraphics[width=0.98\columnwidth]{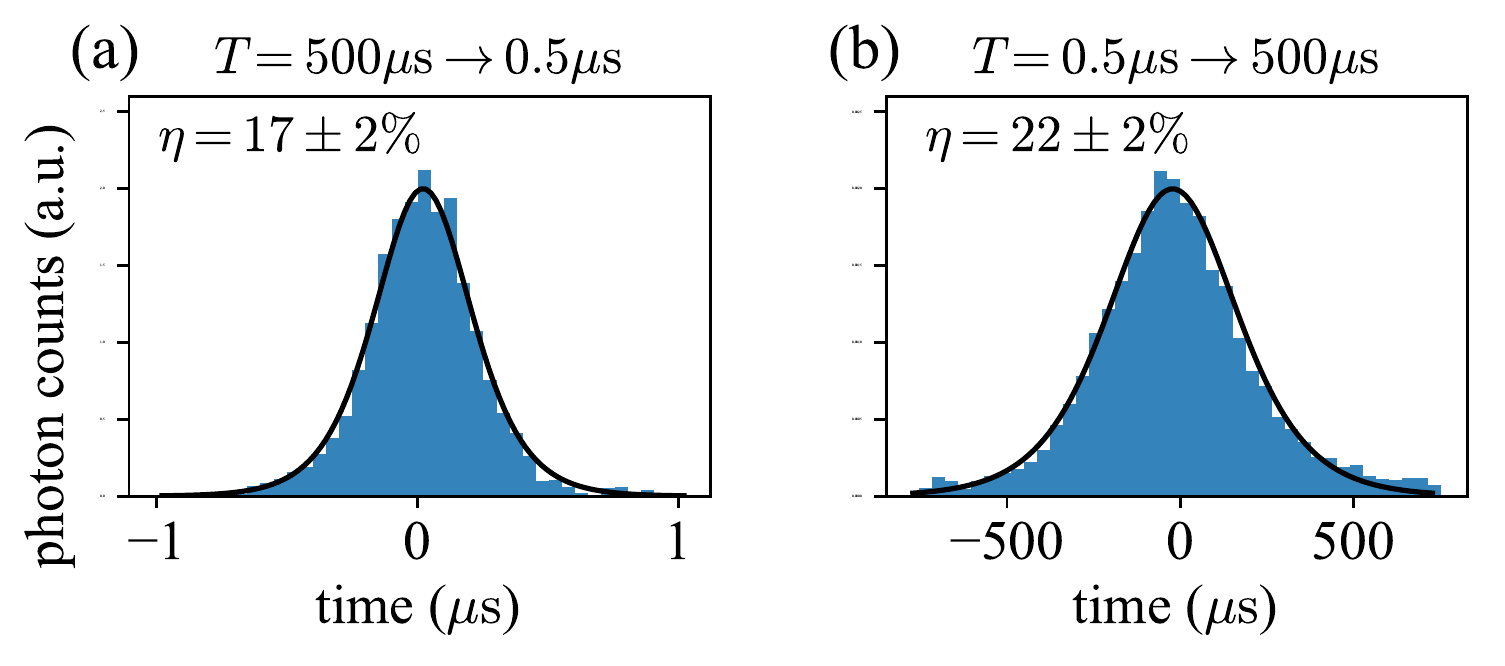}
\caption{Photon conversion: A single photon (experimentally a weak coherent state input) is stored and re-emitted with a different temporal shape with constant efficiency. (a) and (b) illustrate two different photon shape conversions $T=0.5\mu\text{s}\leftrightarrow T=500\mu\text{s}$. The histograms correspond to the detection event of the photon counters and the plain lines correspond to the targeted temporal shapes. In (b) we have subtracted the dark counts of the photon counters. 
}
\label{fig:convert}
\end{figure}

%\part{Shape conversion}
By combining absorption and emission, it is in principle possible to convert between arbitrary shapes. We have realized two different conversions. In the first one, we store a photon of duration $T=500\mu$s and re-emit it with a new duration of $T=0.5\mu$s with an overall efficiency of 17\%. In the second one, we have performed the opposite transformation, from $T=0.5\mu$s to $T=500\mu$s, with an efficiency of 22\%. Figure \ref{fig:convert} displays the two corresponding output shapes measured by photon counting. With a total efficiency of about 20\% and a change of pulse duration by three orders of magnitude, our system outperforms the non-unitary reshaping that can be achieved with any amplitude modulator \cite{Kolchin}. This opens up the possibility to connect quantum devices working at very different time scales.

Although most of our results show a very good agreement between theory and experiment, the efficiency of about 20\% does not match the theoretical value of $(66\%)^2=43\%$, assuming that absorption and emission have the same efficiency. As we have measured an emission efficiency of about 66\%, this tends to indicate that the absorption is a factor of two smaller than expected. This difference remains unexplained and neither the theory nor the experimental imperfections seem to quantitatively explain the observed discrepancy. Nevertheless, except for this point, our efficiency is in principle not fundamentally limited, and ideally a unitary emission/absorption/transformation is achievable with a non-lossy cavity and a higher cooperativity, as explained above.

%\part{Conclusion}
In conclusion, we have shown two main achievements. First, we developed a comprehensive description of light-matter interfaces using CQED. This allowed us to extend the operating parameters, especially the detuning of the control laser and the cavity, over a wide range into a regime that was not explored before. For large detunings, we found it mandatory to compensate the photon phase chirp that stems from the in this case large time-dependent control-laser intensity. We also identified efficiency limitations from destructive interference of emission pathways as well as decoherence issues from spontaneous emission and optical pumping. It is worth noting that the drawn results apply formally to any CQED system, and that most of the observed effects are relevant for other physical platforms \cite{Kurpiers2018}. 

Second, we used this understanding to demonstrate an unprecedented level of control of the temporal shape of a single photon in absorption and emission, and thus transformation. These are crucial capabilities in many quantum information protocols involving single-photon states. For instance, it allows to achieve a high level of indistinguishability between multiple systems and therefore sustains the scalability of single-photon based quantum architectures. Of course, all the discovered features can immediately be transferred to single photons carrying polarization qubits.

Last but not least, the extended parameter regime with the possibility of using the cavity far-detuned from the atomic resonance will be an important tool for coupling individual atoms from a multi-atom quantum register to the same cavity \cite{Enk97}.\\[0.2cm]

%\begin{acknowledgments}
%\part{Acknowledgments}
We thank L. Giannelli and S. Ritter for discussions. This work was supported by the Bundesministerium für Bildung und Forschung via the Verbund Q.Link.X (16KIS0870), by the Deutsche Forschungsgemeinschaft under Germany’s Excellence Strategy – EXC-2111 – 390814868, and by the European Union’s Horizon 2020 research and innovation programme via the project Quantum Internet Alliance (QIA, GA No. 820445).

%\end{acknowledgments}

\bibliographystyle{plain}

\let\oldaddcontentsline\addcontentsline% Store \addcontentsline
\renewcommand{\addcontentsline}[3]{}% Make \addcontentsline a no-op

\let\addcontentsline\oldaddcontentsline% Restore \addcontentsline

%\end{document}

%%%%%%%%%%%%%%%% SUPPLEMENTARY

\pagebreak

\onecolumngrid
\begin{center}
  \textbf{\large Supplemental Material: \\ Deterministic Shaping and Reshaping of Single-Photon Temporal Wave Functions}\\[.2cm]
  O. Morin,  M. Körber, S. Langenfeld, and G. Rempe\\[.1cm]
  {\itshape ${}^1$Max-Planck-Institut f\"{u}r Quantenoptik, Hans-Kopfermann-Strasse 1, 85748 Garching, Germany\\}
(Dated: \today)\\[0.5cm]
\end{center}

\begin{adjustwidth}{55pt}{55pt} \small
In this additional document, we provide details necessary to carry out the results presented in the main text. We first provide the derivation of the main equations. We then discuss some more specific aspects of our results. Eventually, we provide details about the homodyne measurements.\\[0.5cm]
\end{adjustwidth}

\twocolumngrid

\setcounter{equation}{0}
\setcounter{figure}{0}
\setcounter{table}{0}
\setcounter{page}{1}
%\renewcommand{\theequation}{S\arabic{equation}}
%\renewcommand{\thefigure}{S\arabic{figure}}
%\renewcommand{\bibnumfmt}[1]{[S#1]}
%\renewcommand{\citenumfont}[1]{S#1}

%\maketitle

\tableofcontents

\def \cg {[?]\xspace}

\newcommand{\mm}[1]{\ensuremath{#1}\xspace} 
%%% Names of used constants!
\def \DefCooperativity {\mm{ C = \frac{g^2}{2\kappa \gamma}}}
% cavity field defs
\def \dE {\dot{\mathcal{E}}}
\def \E {\mathcal{E}}
\def \sEin {\mm{\mathcal{E}_\text{in}} }
\def \Ein {\mm{\mathcal{E}_\text{in}} }
\def \EinTilde {\mm{\tilde{\mathcal{E}}_\text{in}} }
\def \Eini		{\mm{\mathcal{E}_\text{in}(t')}}
\def \Einrev 	{\mathcal{E}_\text{in}(-t)}
\def \Eout 		{\mm{\mathcal{E}_\text{out}} }
\def \Eoutrev 	{\mathcal{E}^*_\text{out}(-t)}
% gorshkov states
\def \hS 		{{S}}
\def \hP 		{{P}}
\def \dS 		{\dot{\hS}}
\def \dP		{\dot{\hP}}
\def \sEffEsc   {\mm{\sEff_\text{esc}}}
% photon detuning definitions
\def \aSpds    {single-photon detunings\xspace}
\def \sSpdi   {\mm{\Delta_i} }
\def \aSpd    {single-photon detuning\xspace}
\def \sSpd   {\mm{\Delta} }
\def \sOmegaC {\mm{\omega_c} }
\def \aTpd    {two-photon detuning\xspace}
\def \sTpd		{\mm{\delta} }
% Rabi frequency definitions
\def \sRabiC {\mm{\Omega} }
\def \sRabiCcc {\mm{\Omega^*} }
\def \sRabiCi {\mm{\Omega_{i}} }
\def \sRabiCOne {\mm{\Omega_{1}} }
\def \sRabiCN {\mm{\Omega_{N}} }
\def \sRabiCicc {\mm{\Omega_{i}^*} }
% used constants
\def \constOne {a}
\def \constTwo {b}
\def \constK {K}
\def \constL {L}
% decay constants
\def \sADecay {\mm{\gamma}}
\def \sCDecay {\mm{\kappa}}
\def \sEff 	 {\ensuremath{\eta}\xspace}
\def \sEffEsc   {\mm{\sEff_\text{esc}}}
% math
\def \iu {\mm{i}}
%%%
%%% state commands
\renewcommand{\F}[1]{\mm{F{=}#1}}
\newcommand{\FOne}{\F{1}}
\newcommand{\FTwo}{\F{2}}
\newcommand{\FP}[1]{\mm{F'{=}#1}}
\newcommand{\FPOne}{\FP{1}}
\newcommand{\FPTwo}{\FP{2}}
\renewcommand{\FPrime}[1]{\left| F'=#1 \right\rangle}
\renewcommand{\Fmf}[2]{\mm{\left| F{=}#1, m_F{=}#2 \right\rangle}}
\newcommand{\DTwoFmf}[2]{\mm{\left| \RbDTwoS, F{=}#1, m_F{=}#2 \right\rangle}}
\newcommand{\GSFmf}[2]{\mm{\left| \RbGS, F{=}#1, m_F{=}#2 \right\rangle}}
%%% Specific Rubidium states
\def \Rb {\mm{^{\text{87}}\text{Rb }}}
\def \RbGS {\mm{ 5^2S_{1\slash2} }}
\def \RbDOneS {\mm{ 5^2P_{1\slash2}}}
\def \RbDTwoS {\mm{ 5^2P_{3\slash2}}}
\def \DOne {\mm{\text{D}_{1}}}
\def \DTwo {\mm{\text{D}_{2}}}

\renewcommand{\Im}{\operatorname{Im}}
\renewcommand{\Re}{\operatorname{Re}}

\section{Comprehensive model}

As summarized in \cite{Luigi}, various approaches have been proposed to describe the coherent absorption/emission of a single photon from an atomic system coupled to an optical cavity via a two-photon Raman interaction. The solution that considers any arbitrary single-photon detuning was proposed by \cite{Gorshkov}. In the following, we adopt this previous work by considering an imperfect cavity and adding more excited levels to the considered $\Lambda$-scheme. After simplifications with reasonable assumptions, we end up with formally the same system of differential equations as \cite{Gorshkov} and therefore solve the system the same way.
Eventually, we find a new expression for the efficiency (absorption/emission) and a new relation between the temporal mode of the photon and the temporal shape of the control field.

\subsection{The original model}

Remarkably, although in \cite{Gorshkov} the number of atoms is assumed to be large, an assumption which is obviously not possible for our system, they end up with the same differential equation system as \cite{Luigi,Dilley} which consider a single atom in the framework of the input-output formalism \cite{Dalibard}.

Within the Hilbert space $\mathcal{H}_\text{atom}\otimes\mathcal{H}_\text{cavity}\otimes\mathcal{H}_\text{in}\otimes\mathcal{H}_\text{out}$, the model is restricted to the following set of product states $\ket{g,1,0,0}$, $\ket{g,0,1,0}$, $\ket{g,0,0,1}$, $\ket{e,0,0,0}$, $\ket{s,0,0,0}$, each respectively associated to the probability amplitudes $\mathcal{E},\mathcal{E}_\text{in},\mathcal{E}_\text{out},P,S$. For the atom, $\ket{g}$ is the ground state, $\ket{e}$ the excited state and $\ket{s}$ the storage state. Those three levels constitute the $\Lambda$-system.

We will use the notations of the original derivation by \cite{Gorshkov} so that the reader can easily refer to it. One can nevertheless refer to the following table of notations
\begin{equation*}
    \begin{array}{ccccc}
	%	\cite{Dilley} & & \cite{Gorshkov}& & \cite{Luigi}\\
	[2] & & [1] & & [3]\\
	c_g & \leftrightarrow & \E & \leftrightarrow &c\\
	c_x & \leftrightarrow & P & \leftrightarrow &e\\
	c_e & \leftrightarrow & S & \leftrightarrow &r
    \end{array}.
\end{equation*}

We start with the input-output relation of the quantized optical field for the cavity:
\begin{equation}
\Eout = \sqrt{ 2 \kappa } \E - \Ein \label{eqn:g_inputoutput_gdl}
\end{equation}
with $2\kappa$ being the cavity field decay rate.

\begin{figure}[!t]
\includegraphics[width=0.8\columnwidth]{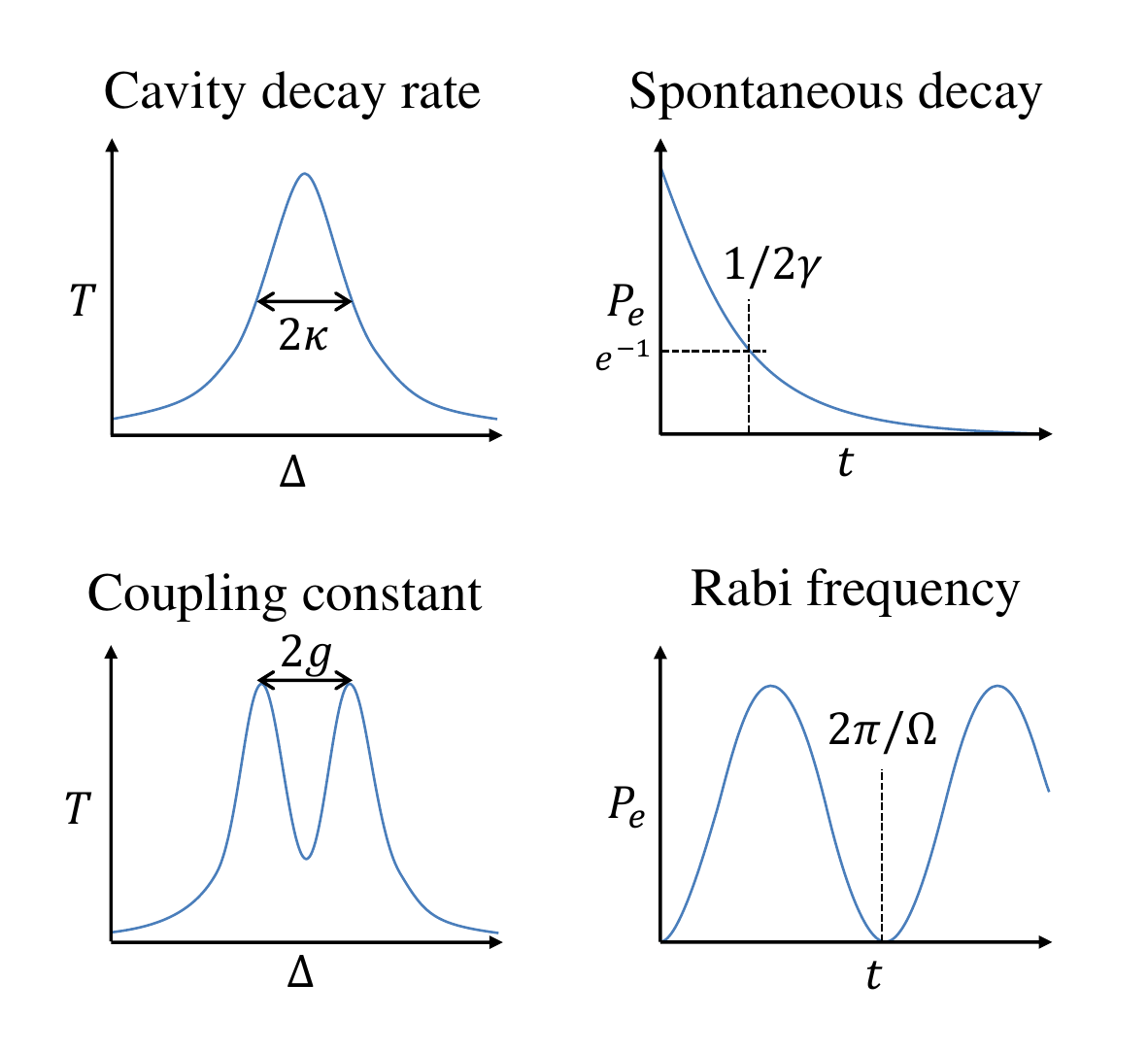}
\caption{Convention used for the definition of the cavity quantum electrodynamics parameters. We relate all four to the directly measurable quantities. The cavity field decay rate $2\kappa$ corresponds to the full-width-half-maximum of the cavity spectrum (here in transmission). The atomic polarization decay rate, $\gamma$, is related to the exponential decay the population of the excited state of the atom. The light-matter coupling constant, $g$ (also called CQED coupling constant), relates to the splitting of the normal mode for a cavity coupled to an atom (on resonance with the atomic transition). The Rabi frequency corresponds to the angular frequency of the atomic population oscillations (when driven by an external coherent field).  }
\label{fig:convention}
\end{figure}

Here, $\sADecay$ is the atomic polarization decay rate, $g$ the cavity coupling rate and $\Omega$ the Rabi frequency of the control field (treated as classical). Note that our definition of Rabi frequency is different from \cite{Gorshkov} but the same as \cite{Dilley}. 
Hence, one can write the equations of motion within the rotating frame:
\begin{align}
\dE &= - \sCDecay \E + \iu g P + \sqrt{2 \sCDecay} \Ein, \label{eqn:g_field_dgl} \\
\dP &= - \left(\gamma + \iu \sSpd \right) \hP + i g \E + i \tfrac{1}{2}\sRabiC \hS, \label{eqn:g_estate_dgl} \\
\dS &=  i \tfrac{1}{2}\sRabiCcc \hP. \label{eqn:g_sstate_dgl} 
\end{align}
In \Fig \ref{fig:convention}, we specify the definition of the different parameters.

\subsection{Imperfect cavity}

The optical cavity is not perfect, and it is necessary to model it with two decay channels: one to the output field, $\kappa_c$, and a second for optical losses, $\kappa_l$. This later including the transmission through the high-reflectivity mirror and the intra-cavity losses. Then, we introduce the escape efficiency
\begin{equation}
\sEffEsc = \frac{\sCDecay_c}{\kappa_c+\kappa_l}
\end{equation}
which can be interpreted as the ratio of the probability of the cavity field to escape through the output field, proportional to $\kappa_c$, to the probability to escape from the cavity, including the output field and the losses channel $\kappa_c+\kappa_l$.

Hence, the input-output relation (\ref{eqn:g_inputoutput_gdl}) and the equation of motion (\ref{eqn:g_field_dgl}) for the cavity field become
\begin{align}
\Eout &= \sqrt{ 2 \sEffEsc \kappa } \E - \Ein, \label{eq:g_inputoutput_gdl2}\\ 
\dE &= - \sCDecay \E + \iu g \hat{P} + \sqrt{2 \sEffEsc \sCDecay} \Ein. 
\label{eqn:gorshkovinout}
\end{align}
As we will see, this change simply adds a prefactor to the efficiencies.

\subsection{3 excited states}

\begin{figure}[!t]
\includegraphics[width=\columnwidth]{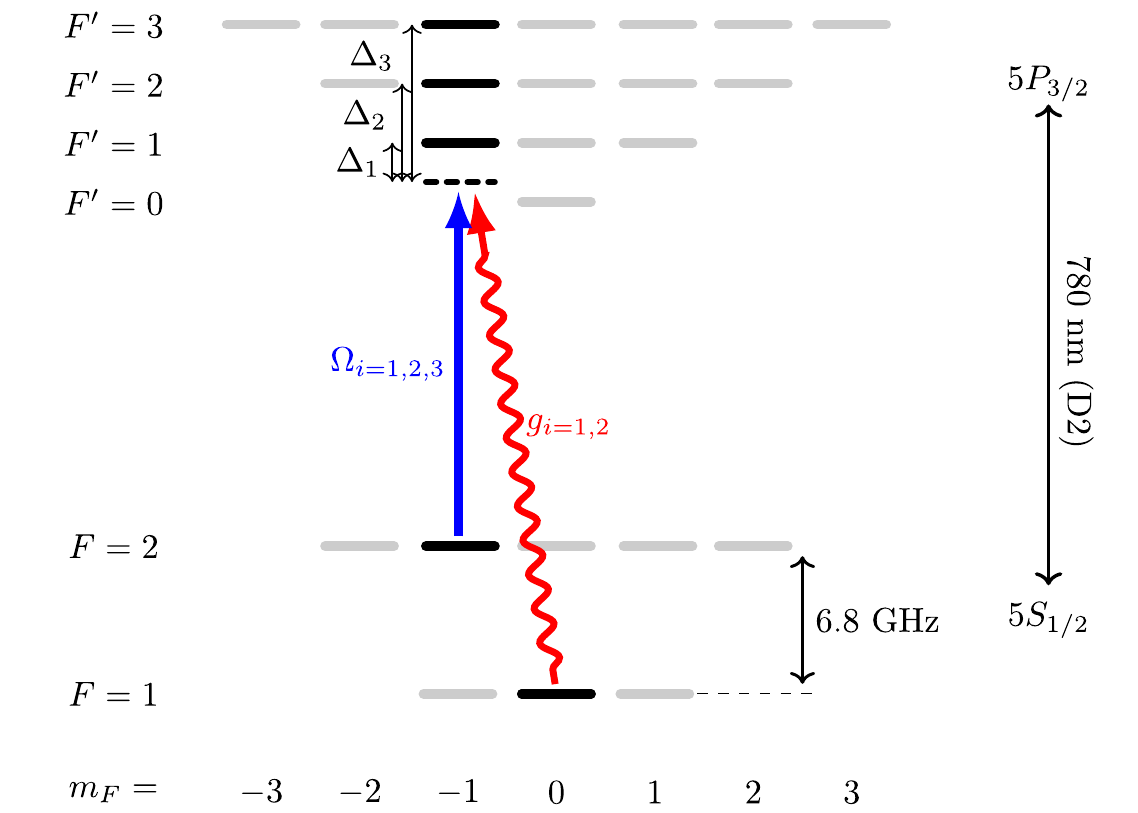}
\caption{Atomic level structure of $^{87}$Rb used for our model. In black, the levels considered for the model. In red, the cavity field which couples $F=1$ to the excited states $F'=1$ and $F'=2$ with different strength $g_{i=1,2}$. In blue, the control field coupling $F=2$ to $F'=1,2,3$ with the Rabi frequencies $\Omega_{i=1,2,3}$.}
\label{fig:levels_sup}
\end{figure}

In order to describe our experiment completely, we introduce $P_1,\ P_2,\ P_3$ for the excited states $F'=1,\ 2, \ 3$ (note that we ignore $F'=0$ as it is not coupled to any of the ground states for the considered optical fields). We also define the associated cavity coupling rates, $g_i$, cooperativities, $C_i$, and Rabi frequencies, $\Omega_i$,  as 
\begin{equation*}
\sRabiCi = c_{si} \sRabiC,
\quad g_i=c_{gi}g
,\quad C_i = \frac{g_i^2}{2\sADecay \sCDecay}
\end{equation*}
with $c_{si}$ (resp. $c_{gi}$) being the Clebsch-Gordan coefficients for the transition between the state $\ket{s}$ (resp. $\ket{g}$) and the excited states $\ket{e_i}$.
We then obtain the following system of linear differential equations:
\begin{align}
\dE &= - \sCDecay \E +  \iu g_1 P_1 + \iu g_2 P_2 + \sqrt{2 \sEffEsc \sCDecay} \Ein \label{eqn:e_field_dgl} \\
\dP_1 &= - \left(\gamma + \iu \sSpd_1 \right) P_1 + i g_1 \E + i \tfrac{1}{2}\sRabiC_1 S\\
\dP_2 &= - \left(\gamma + \iu \sSpd_2 \right) P_2 + i g_2 \E + i \tfrac{1}{2}\sRabiC_2 S \\
\dP_3 &= - \left(\gamma + \iu \sSpd_3 \right) P_3 + i \tfrac{1}{2}\sRabiC_3 \hS\\
\dS &=  i \tfrac{1}{2}\sRabiCcc_1 P_1 + i\tfrac{1}{2}\sRabiCcc_2 P_2 + i\tfrac{1}{2}\sRabiCcc_3 \hP_3.
\end{align}

\newcommand{\rchi}{\eta_\text{esc}}
%
% Adiabatic Elimination 
%
\subsection{Analytical adiabatic solutions}
\label{ssub:assumptions}
To find an analytical solution, we apply two assumptions analogously to Gorshkov \textit{et al}. \cite{Gorshkov}. 
First, we assume the bad cavity limit
\begin{equation}
\kappa \gg g
\label{eqn:badcavity}
\end{equation}
such that one can adiabatically eliminate the cavity field, i.e. $\dot{\mathcal{E}}\approx0$.
This is obviously a questionable approximation in our case. However, \cite{Luigi} has shown that the analytical solution one obtains remains satisfying as long as one uses a smooth enough shape for $e(t)$.
Second, we assume smoothly varying control Rabi frequencies, $\Omega_i(t)$, and a smooth incoming photon, $\Ein(t)$ (or out-coming photon $\Eout(t)$),  such that one can adiabatically eliminate the $P_i$s, i.e. $\dot{P}_i\approx0$. The equations then simplify as follows
\begin{align}
\E &= \frac{1}{\sCDecay} \left(  \iu g_1 P_1  +\iu g_2 P_2  + \sqrt{2 \sEffEsc \sCDecay} \Ein \right)  \label{eqn:ea_field_dgl}\\
P_1 &=  \frac{1}{\left(\gamma + \iu \sSpd_1 \right)} \left( i g_1 \E + i\tfrac{1}{2} \sRabiC_1 S \right) \label{eq:P1}\\
P_2 &=  \frac{1}{\left(\gamma + \iu \sSpd_2 \right)} \left( i g_2 \E + i\tfrac{1}{2} \sRabiC_2 S \right) \label{eq:P2}\\
P_3 &=  \frac{1}{\left(\gamma + \iu \sSpd_3 \right)} \left(i \tfrac{1}{2}\sRabiC_3 \hS \right) \label{eq:P3}\\
\dS &=  i\tfrac{1}{2} \sRabiCcc_1 \hP_1 + i\tfrac{1}{2}\sRabiCcc_2 \hP_2 + i\tfrac{1}{2}\sRabiCcc_3 \hP_3 . 
\end{align}

In order to simplify the expressions, we introduce the intermediate parameters
\begin{equation}
\constOne_j = { \sADecay \left(1+2C_j \right) + i \sSpdi }\  , \qquad \constTwo = \frac{g_1 g_2 }{\sCDecay} \qquad.
\end{equation}

Therefore, on can solve the previous system and obtain the following differential equation
\begin{equation}
\dot{S}=-K\abs{\Omega}^2S \label{eq:eq_diff_S}
\end{equation}
with the parameter 
\begin{equation}
K=\frac{1}{4}\left[ \frac{c_{s1}^2 \constOne_2 + c_{s2}^2 \constOne_1 - 2c_{s2}c_{s1}\constTwo}
	{\constOne_1 \constOne_2 - \constTwo^2}
	+ \frac{c_{s3}^2}{\constOne_3}\right].
\end{equation}

\Eq (\ref{eq:eq_diff_S}) is a first order linear differential equation with a variable coefficient. We thus get the solution
\begin{equation}
S(t) = \hS(t_0) \exp \left( - \constK \int_{t_0}^t |\Omega(t')|^2 dt'  \right).
\end{equation}

\subsection{Photon emission}
In the case of photon emission, there is no incoming photon, i.e. $\Ein=0$, and we start with a fully populated storage state $\hS(t_0=0)=1$. 
For convenience, we use the notation
\begin{align*}
&h(t) := \int_{t_0}^t |\Omega(t')|^2 dt',\\
&\partial_t h(t) = \abs{\Omega(t)}^2\ .
\end{align*}
Therefore, Equation (\ref{eqn:ea_field_dgl}) becomes
\begin{equation}
\E = \frac{\iu}{\kappa} \left( 
g_1 P_1 + g_2 P_2 \right)
\end{equation}
and Equation (\ref{eq:g_inputoutput_gdl2}) becomes
\begin{equation}
\Eout = \sqrt{\eta_{esc}}\sqrt{2\kappa}\E\ .
\end{equation}

Then, by using Eq. (\ref{eq:P1})-(\ref{eq:P3}), one can find
\begin{equation}
\Eout =\sqrt{\eta_{esc}} L \sRabiC S\ . \label{eq:Eout0}
\end{equation}
with
\begin{equation*}
L = \frac{1}{2}\sqrt{2\gamma  C }
	\frac{ c_{g1} \left(\constOne_2 c_{s1} - c_{s2} \constTwo \right)
		+
		c_{g2} \left(\constOne_1 c_{s2} - c_{s1} \constTwo \right)
	}
	{\constTwo^2-\constOne_1 \constOne_2}.
\end{equation*}
Again, all the properties of the atom-cavity system are integrated in the time-independent parameter $\constL$.

\subsubsection{Photon emission efficiency}
In the next step, we want to calculate the efficiency of the photon emission process. This can be evaluated by integrating the outcoming-mode:
\begin{equation}
\begin{split}
\eta_R 
=& \int_{t_0}^\infty |\Eout|^2 dt  \\
=&  |\constL|^2 \int_{t_0}^\infty \partial_t h(t) \cdot e^{ - 2 \Re \left( \constK \cdot  h(t) \right)} dt \\
=& \frac{- |\constL|^2}{2 \Re(\constK)} \left[ e^{ - 2 \constK h(t \to \infty) } - e^{ - 2 \constK h(t_0) } \right] \ .
\end{split}
\end{equation}
By assuming $2 \constK h(t \to \infty) \rightarrow \infty$, we obtain the efficiency
\begin{equation}
\eta_R \approx  \frac{|\constL|^2}{2 \Re(\constK)}.
\end{equation}

In contrast to \cite{Gorshkov}, the efficiency here depends on the single-photon detuning (contained in the parameters $K$ and $L$).

\subsubsection{Photon shape control}
Here, the mapping between the control Rabi frequency, \sRabiC, and the created photon shape, \Eout, is derived.
We first introduce the temporal shape of the out-coming photon. This latter is normalized and thus 
\begin{align}
e(t) = \sqrt{\eta_R}^{-1} \Eout \ .
\end{align}
We start by using the equation (\ref{eq:Eout0})
\begin{equation}
\int_{0}^t dt'\abs{e(t')}^2=1-\exp(-2\Re(K)h(t)).
\end{equation}
Then taking the time derivative we get
\begin{equation}
\abs{e(t)}^2=2\Re(K)\abs{\Omega(t)}^2\underbrace{\exp(-2\Re(K)h(t))}_{\int_t^\infty\abs{e(t')}^2dt'}
\end{equation}
which can be transformed to
\begin{equation}
\abs{\Omega(t)}=\frac{1}{\sqrt{2\Re{K}}}\frac{\abs{e(t)}}{\sqrt{\int_t^\infty\abs{e(t')}^2dt'}}\ . \label{eq:absOmega}
\end{equation}
We now need to calculate the phase of $\Omega$. As equation (\ref{eq:Eout0}) gives
\begin{equation}
\Eout(t) = L\Omega(t)\exp(-Kh(t)),
\end{equation}
we obtain from (\ref{eq:absOmega})
\begin{equation}
\arg(\Eout)=\arg(e)=\arg(L)+\arg(\Omega)-\Im(K)h(t)
\end{equation}
that we can rewrite to
\begin{equation}
\arg(\Omega)=\arg(e)-\arg(L)+\Im(K)h(t)\ .
\end{equation}
First, $L$ only contributes with a time-independent phase term which we omit in the following.
Second, in order to express $\Omega$ as a function of $e(t)$ only, we have to substitute $h(t)$ in the phase term via the equation
\begin{equation}
\begin{split}
\exp(-2\Re(K)h(t))&=1-\int_0^t\abs{e(t')}^2dt'\\
&=\int_t^\infty\abs{e(t')}^2dt'
\end{split}
\end{equation}
such that
\begin{equation}
h(t) = \frac{1}{2\Re(K)}\ln\left(\int_t^\infty\abs{e(t')}^2dt'\right).
\end{equation}
We can now write the Rabi frequency of the control pulse as a function of the desired output photon temporal shape
\begin{multline}
\sRabiC(t) = \frac{e(t)}{ \sqrt{ 2 \Re (\constK) \int_t^\infty |e(t')|^2 dt' }} \\
\exp\left(- \iu \frac{\Im( \constK )}{2\Re(\constK)} \ln \left( \int_t^\infty |e(t')|^2 dt' \right) \right) \ .
\end{multline}

Although this expression is analytically of the same form as the one derived in \cite{Gorshkov}, the new term $K$ is making all the difference.

\subsection{Photon absorption}
\label{sub:photon_absorption}
The storage of an incoming photon can be seen as the time-reserved process of photon absorption as explained in \cite{Gorshkov}.
We can then use the transformations
\begin{align}
\Eout &\rightarrow \Eoutrev = \Ein(t) \\
\Omega(t) &\rightarrow \Omega^*(-t) = \Omega_{0,\text{st}}(t)
\end{align}
assuming $t_0=0$ at the beginning of the retrieval or, symmetrically, at the end of the storage.
The control field Rabi frequency for photon storage can be directly obtained
\begin{multline}
\Omega_\text{st}= \frac{e_\text{in}(t)}{ \sqrt{ 2 \Re (\constK) \int_0^t|e_\text{in}(t')|^2 dt' }}\\ 
\exp\left(\iu \frac{\Im(\constK)}{2\Re (\constK)} \ln \left( \int_0^t|e_\text{in}(t')|^2 dt' \right) \right).
\end{multline}
It directly follows from the time-reversal argument that the absorption efficiency equals the emission efficiency for an incoming photon which is perfectly coupled to the cavity mode.

\section{Spontaneous decay and incoherent photon emission}

The analytical solution that we have derived assumed that any spontaneous decay systematically ends the photon emission/absorption process by removing the excitation from the system. This is true for a system with many atoms as the storage is carried by a spin-wave excitation, i.e. any spontaneous decay will destroy the coherence of the spin-wave. For a single atom, the situation is different as there are no coherences to destroy. Hence, after scattering a photon, the atomic population has a non-vanishing probability to decay to a state within the $F=2$ ground state manifold.
All states within this manifold are coupled by the control pulse to at least one excited state which forms a $\Lambda$-system with the cavity mode.
Hence, if the control beam is still on, a new attempt of photon generation starts.

\begin{figure}[!ht]
\includegraphics[width=\columnwidth]{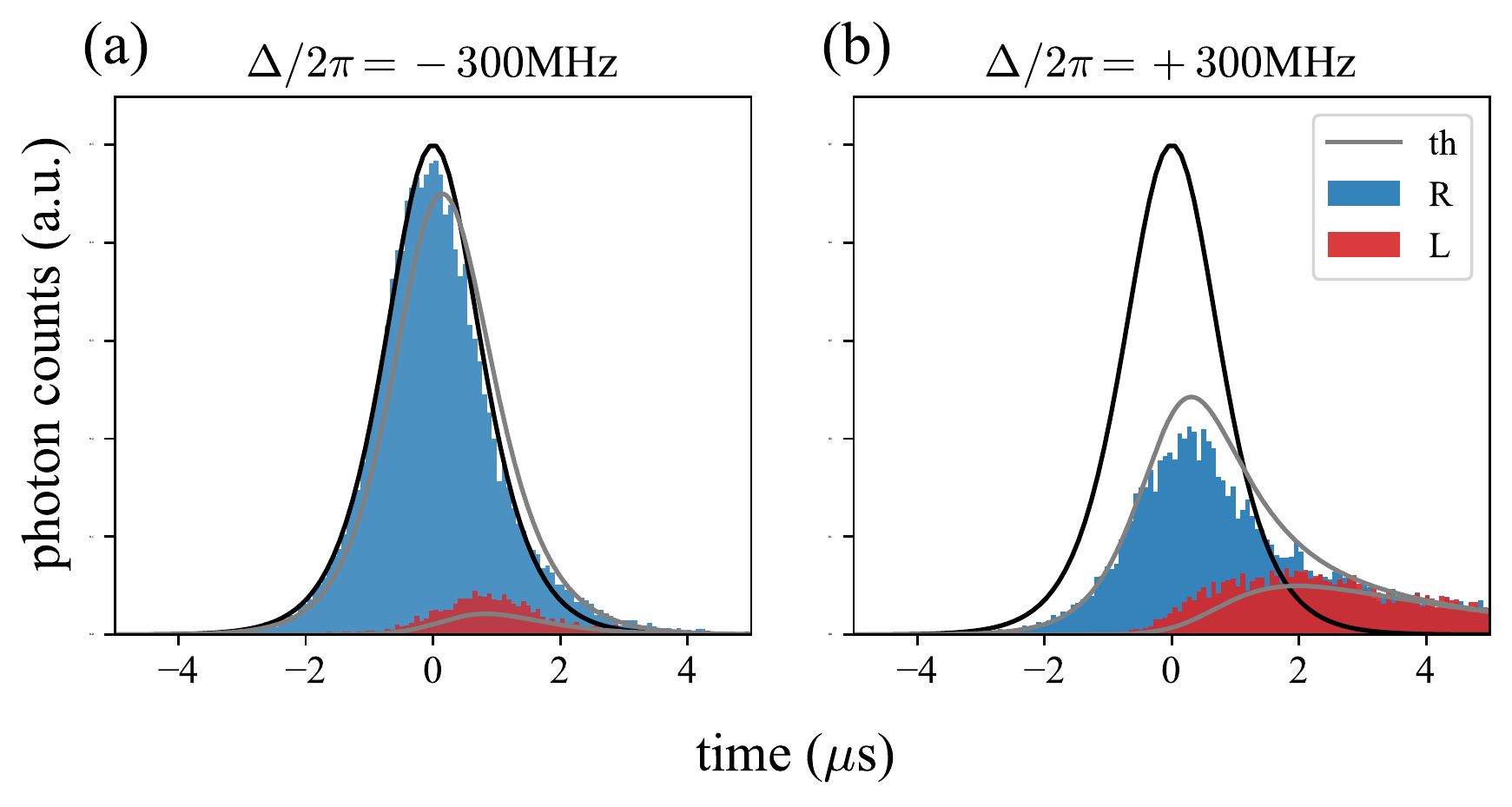}
\caption{Spontaneous decay at large detuning. Photon shape histogram for two single-photon detunings $\Delta/2\pi=\pm300$MHz recorded by photon counting for the two circular polarizations. The grey lines indicate the shape obtained by simulation. }
\label{fig:shape_and scat}
\end{figure}

Figure \ref{fig:shape_and scat} reports the photon shape presented in the main text in the case of large single-photon detuning with respect to $F'=1$. We essentially notice here the changes in the photon shape and some detection events on the orthogonal polarization. This is the main consequence of spontaneous decay. The grey lines correspond to the simulation of the system including those different photon emission channels.

\subsection{Numerical simulations}
To quantify the impact of incoherent processes started by spontaneous atomic decay, we employed the \textit{Quantum Toolbox in Phyton}(QuTiP) to numerically integrate the dynamics of the system.
To this end, the system is simulated by implementing all ground states, as well as all $13$ relevant states of the $D_2$ line, which are coupled by the control pulse to one of the states in $F=2$.
Additionally, both the $\sigma^+$- and $\sigma^-$-modes of the cavity are added to the system.
Furthermore, for every cavity mode, two additional states for the out-coupled and the lost population are added.

A unitary Hamiltonian contains the $13$ coupling terms associated with the coupling of the control pulse to the excited states, and the $12$ coupling terms corresponding to the coupling of the two cavity modes with the $F=1$ ground state manifold.
The out-coupling and losses from the cavity mode are described with collapse operators with decay rates $\kappa_c$ and $\kappa_l$, respectively.
Analogously, for every atomic decay channel, a collapse operator term is added with the atomic decay rate, $\gamma$, weighted by normalized Clebsch-Gordan coefficients.

To integrate the system, the unitary Hamiltonian, along with the collapse operators, are passed to the master equation solver of the QuTiP package.

The produced photon is determined by the time-evolution of the expectation value of the states associated with the out-coupling of the cavity-mode.
Note, that this model only allows for the simulation of the photon emission process, but not for the photon absorption.

\begin{figure}[!t]
\includegraphics[width=0.49\columnwidth]{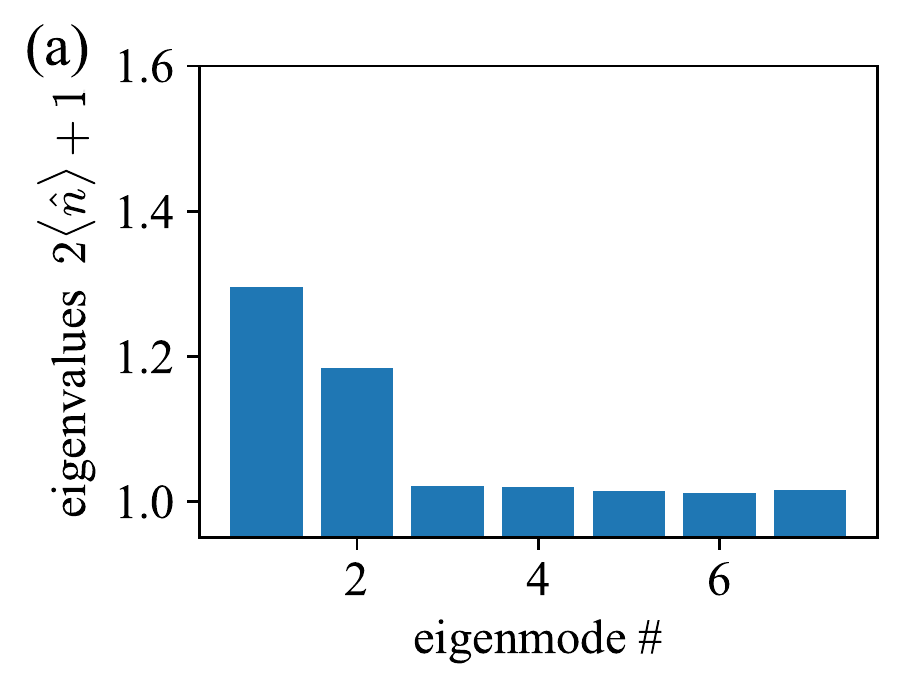}
\includegraphics[width=0.49\columnwidth]{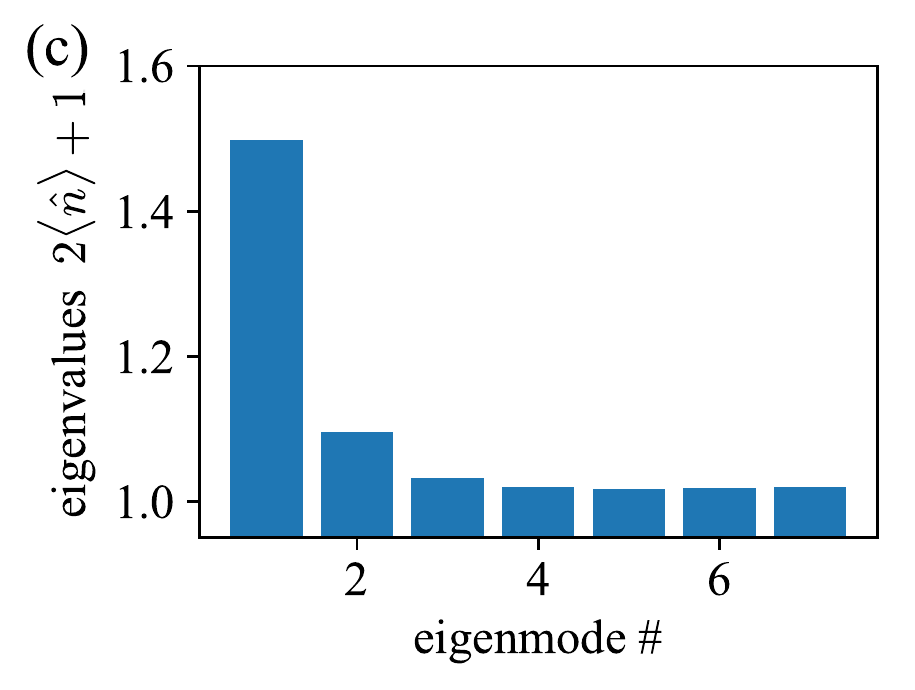}
\includegraphics[width=0.49\columnwidth]{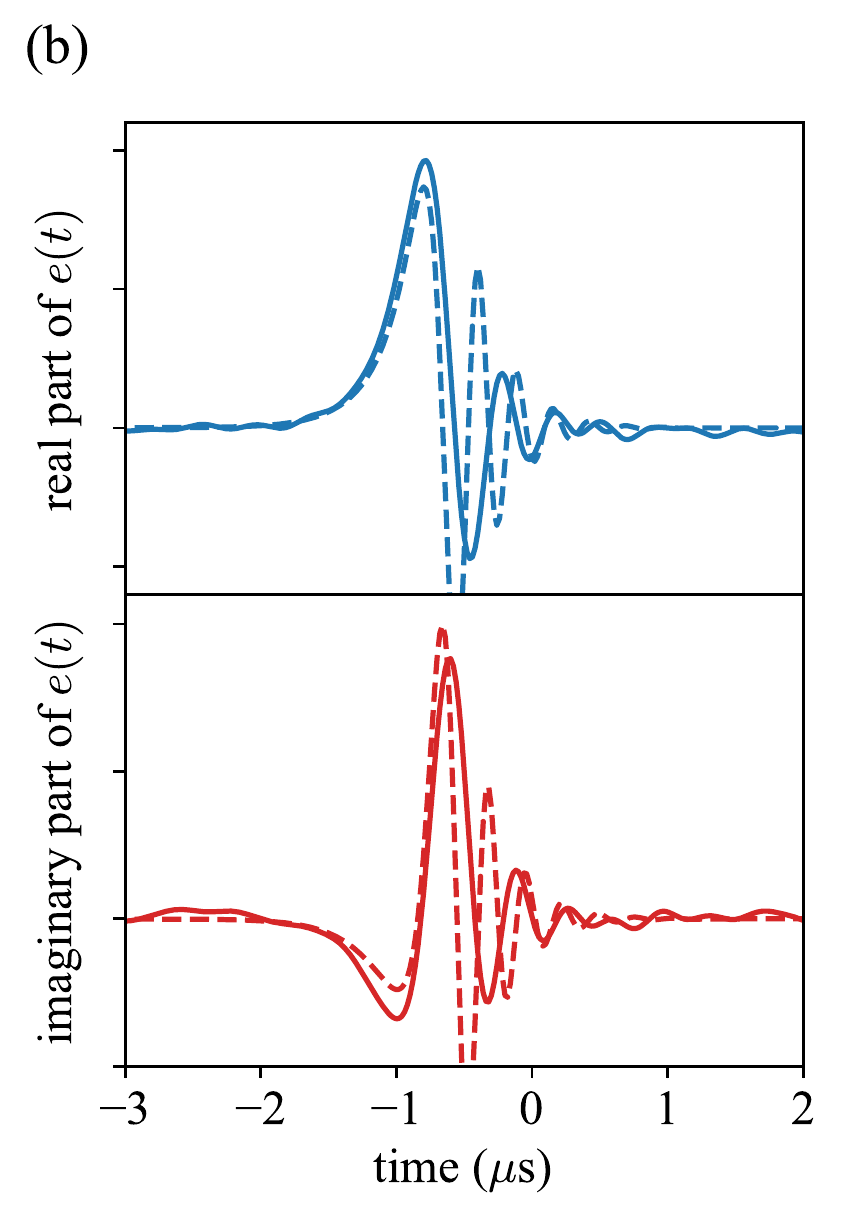}
\includegraphics[width=0.49\columnwidth]{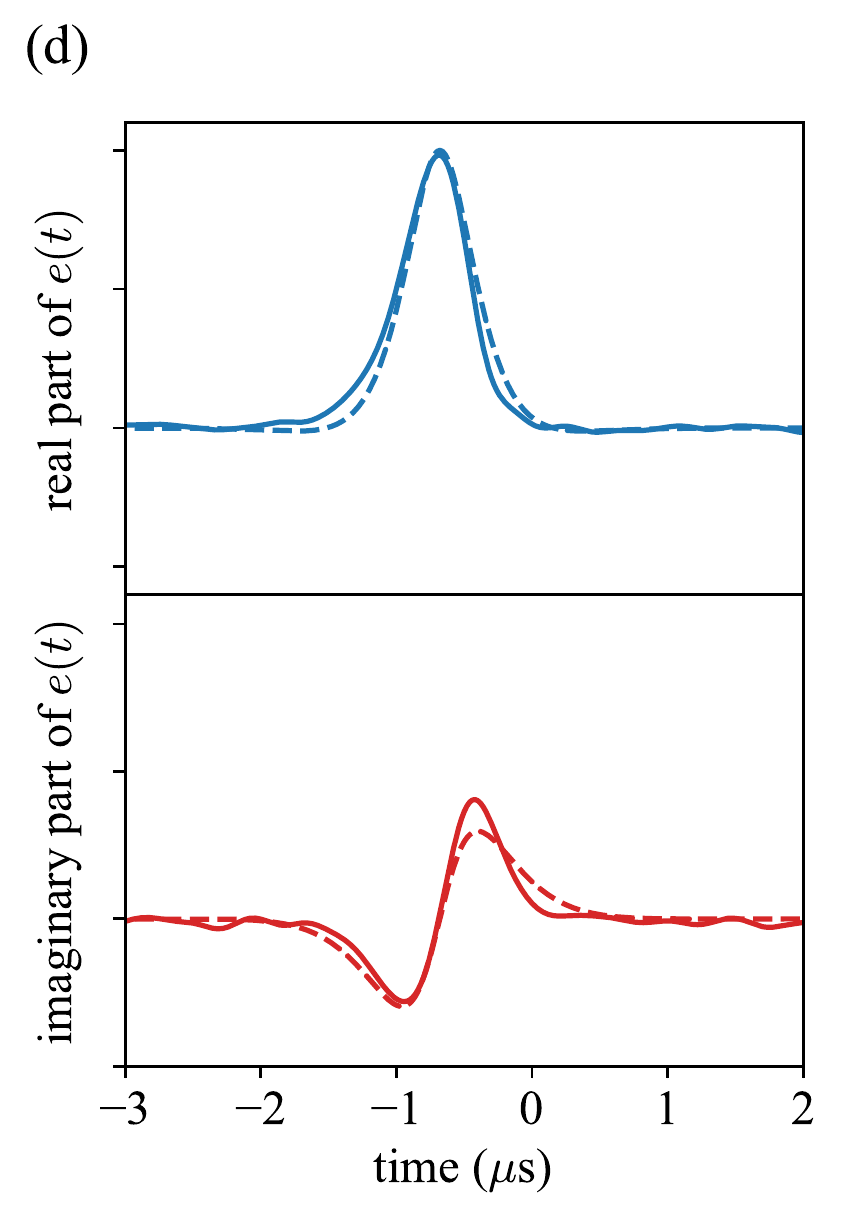}
\caption{Temporal mode measurement for a single-photon detuning of $\Delta/2\pi = -40$MHz. Similarly to the result presented in the main text we have on the left the single photon generated without applying any phase compensation on the control field and on the right the case when the control pulse includes the phase compensation. The histograms (a) and (c) correspond to the measured eigenvalues (which directly relate to the average photon numbers). The plots (b) and (d) correspond to the reconstructed real and imaginary part of the single photon temporal shape. All axes have the same scale.}
\label{fig:temporal_shapeS}
\end{figure}

\section{Homodyning}
Here we provide details and discussions about the temporal mode measured with a homodyne detection\cite{Morin2019}.

\subsection{Temporal mode expansion}
Given a set of quadrature measurements $x_k(t)$, one can compute the auto-correlation function $\mean{x_k(t)x_k(t')}{k}$. Then, we compute its  eigenfunctions and their associated eigenvalues. The set of eigenfunctions that we obtain is orthogonalized, i.e. $\int_\mathbb{R} f_i(t)f_j(t)dt=\delta_{ij}$.
We take a set of measurements of the vacuum state by simply blocking the signal such that we have a reference. Normalized by the vacuum state, each eigenvalue $\kappa_i$, associated to the eigenfunction $f_i$, corresponds to the energy in the mode defined by this eigenfunction. Therefore we have the equality $\kappa_i = 2n_i+1$, where $n_i$ is the average photon number in the mode $f_i$.

Assuming that the signal field contains only a mixture of single-photon and vacuum states and that the single-photon state is in a pure temporal mode, one can obtain at most two eigenvalues above one. (More eigenvalues would indicate that there is more than one temporal mode not being a vacuum state.) The temporal mode of the single photon can be reconstructed via
\begin{equation}
f(t)=\frac{1}{\sqrt{n_1+n_2}}\left(\sqrt{n_1}f_1(t)+i\sqrt{n_2}f_2(t)\right) \ .
\end{equation}

Note that if we only have one eigenvalue then only one mode counts.

Hence, from this, one can compute the phase of the temporal mode $f$ to be
\begin{equation}
\phi(t)=\arctan\frac{\sqrt{n_2}f_2(t)}{\sqrt{n_1}f_1(t)}.
\end{equation}

This latter raises the question of the sign ambiguity of the phase. Indeed, there is no way to determine which mode is $f_1$ and which one is $f_2$. However, as explained in \cite{Morin2019}, this ambiguity can be lifted with an additional measurement with a detuned local oscillator. Nevertheless, in our case this ambiguity is not a relevant issue as we have a very good agreement between theory and experiment.

\subsection{Efficiency}
In the following, we provide an exhaustive list of optical losses present in our experimental setups:
\begin{itemize}
    \item $0.74(5)$ for the atom preparation. The microwave driving is not fast enough to compensate broadening induced by the fluctuating magnetic field. Therefore, only a part of the population ends up in the state $\ket{F=2,m_F=-1}$ whereas the rest remains in the initial state $\ket{F=1,m_F=0}$.
    \item 0.66 for the photon production. This number is given by the theory and has been confirmed by the measurement with the single photon detector (SPD).
    \item $0.90(1)$ for the fiber coupling, which includes the mode matching between the cavity and the fiber, the fiber coupler losses at both ends, and the fiber-losses themselves.
    
    \item $0.970(5)$ for the isolator.
    \item $0.88(1)$ for the mode matching between the local oscillator and the signal mode. This number is equal to the visibility squared (here 0.94\%).
    \item $0.89(5)$ for the efficiency of the photodiode.
    \item 0.98 for the contribution of the electronic noise (which corresponds to 17dB between the signal shot noise and the electronic noise).
    \item $0.90(1)$ basic optical components losses like mirrors, wave plates and filters.
\end{itemize}

The overall efficiency is then about $0.30(5)$ which is in a good agreement with the homodyne measurement for which an average photon number of 0.3 has been measured. 

This is important because it confirms that all the photons are in a single mode (real or complex). As mentioned at the end of the main text, the efficiency in absorption is significantly smaller than the efficiency in emission (this latter being in a good agreement with the efficiency predicted with the model). It is clear that the absorption case is more sensitive to experimental imperfections as the incoming photon and the control field have to match each other, whereas in emission, the cavity and output field are adapting to the control field. The homodyne measurement can measure the temporal shape of the photon but remains with a limited bandwidth whereas SPDs can detect a large range of frequencies. By measuring the same efficiency than the SPD, one rules out the possibility that the system emits statistically into different modes. This could have been an explanation of the observed reduced absorption efficiency.

\bibliographystyle{plain}

\begin{thebibliography}{10}


\bibitem{Knill2001} E. Knill, R. Laflamme, G. J. Milburn, \ShowTitle{A scheme for efficient quantum computation with linear optics, }Nature \textbf{409}, 46–52 (2001).

\bibitem{Minzioni2019}P. Minzioni \textit{et al.}, Roadmap on all-optical processing, J. Opt. \textbf{21}, 063001 (2019).
%https://doi.org/10.1088/2040-8986/ab0e66

\bibitem{Kimble2008} H. J. Kimble, \ShowTitle{The quantum internet, }Nature \textbf{453}, 1023–1030 (2008).


\bibitem{Wehner2018} S. Wehner, D. Elkouss and R. Hanson, \ShowTitle{Quantum internet: A vision for the road ahead, }Science \textbf{362} eaam9288 (2018).


\bibitem{Rohde2005} P. P. Rohde, T. C. Ralph, and M.l A. Nielsen, \ShowTitle{Optimal photons for quantum-information processing, }Phys. Rev. A \textbf{72}, 052332 (2005).


\bibitem{Raymer} M. G. Raymer, and K. Srinivasan, \ShowTitle{Manipulating the color and shape of single photons, }Physics Today \textbf{65}, 11, 32 (2012).

\bibitem{Keller2004} M. Keller, B. Lange, K. Hayasaka, W. Lange,  and H. Walther, \ShowTitle{Continuous generation of single photons with controlled waveform in an ion-trap cavity system, }Nature \textbf{431}, 1075–1078 (2004).

\bibitem{Eisaman2004} M. D. Eisaman, L. Childress, A. André, F. Massou, A. S. Zibrov, and M. D. Lukin, \ShowTitle{Shaping Quantum Pulses of Light Via Coherent Atomic Memory, }Phys. Rev. Lett. \textbf{93}, 233602 (2004).

\bibitem{Eisaman2005} M. D. Eisaman, A. André, F. Massou, M. Fleischhauer, A. S. Zibrov, and M. D. Lukin, \ShowTitle{Electromagnetically induced transparency with tunable single-photon pulses, } Nature \textbf{438}, 837–841 (2005).

\bibitem{Balic2005} V. Bali{\'c}, D. A. Braje, P. Kolchin, G.Y. Yin, and S. E. Harris, \ShowTitle{Generation of Paired Photons with Controllable Waveforms, }Phys. Rev. Lett. \textbf{94}, 183601 (2005).

\bibitem{Kolchin} P. Kolchin, C. Belthangady, S. Du, G. Y. Yin, and S. E. Harris, \ShowTitle{Electro-Optic Modulation of Single Photons, }Phys. Rev. Lett. \textbf{101}, 103601 (2008).


\bibitem{Specht2009} H. P. Specht, J. Bochmann, M. M\"{u}cke, B. Weber, E. Figueroa, D. L. Moehring, and G. Rempe, \ShowTitle{Phase shaping of single-photon wave packets, }Nature Photon. \textbf{3}, 469-472 (2009).


\bibitem{Kielpinski2011}D. Kielpinski, J. F. Corney, and H. M. Wiseman, Quantum OpticalWaveform Conversion, Phys. Rev. Lett. \textbf{106}, 130501 (2011).

\bibitem{McKinstrie2012} C. J. McKinstrie, L. Mejling, M. G. Raymer, and K. Rottwitt, Quantum-state-preserving optical frequency conversion and pulse reshaping by four-wave mixing, Phys. Rev. A \textbf{85}, 053829 (2012).


\bibitem{Zhou2012}S. Zhou, S. Zhang, C. Liu, J. F. Chen, J. Wen, M. M. T. Loy, G. K. L. Wong, and S. Du, \ShowTitle{Optimal storage and retrieval of single-photon waveforms, }Opt. Express \textbf{20}, 24124-24131 (2012).


\bibitem{NisbetJones2011} P. B. R. Nisbet-Jones, J. Dilley, D. Ljunggren, and A. Kuhn, \ShowTitle{Highly efficient source for indistinguishable single photons of controlled shape, }New J. Phys. \textbf{13} 103036 (2011).

\bibitem{NisbetJones2013} P. B. R. Nisbet-Jones, J. Dilley, A. Holleczek, O. Barter, and A. Kuhn, \ShowTitle{Photonic qubits, qutrits and ququads accurately prepared and delivered on demand, }New J. Phys. \textbf{15}, 053007 (2013).

\bibitem{Farrera2016} P. Farrera, G. Heinze, B. Albrecht, M. Ho, M. Ch\'avez, C. Teo, N. Sangouard, and H. de Riedmatten, \ShowTitle{Generation of single photons with highly tunable wave shape from a cold atomic ensemble, }Nature Comm. \textbf{7}, 13556 (2016).

\bibitem{Matsuda2016} N. Matsuda, Deterministic reshaping of single-photon spectra using cross-phase modulation, Sci. Adv. \textbf{2}, e1501223 (2016).


\bibitem{Fisher2016} K. A. G. Fisher, D. G. England, J.-P. W. MacLean, P. J. Bustard, K. J. Resch, and B. J. Sussman, Frequency and bandwidth conversion of single photons in a room-temperature diamond quantum memory, Nature Comm. \textbf{7}, 11200 (2016).


\bibitem{Karpinski2017} M. Karpiński, M. Jachura, L. J. Wright, and B. J. Smith, Bandwidth manipulation of quantum light by an electro-optic time lens, Nature Photon. \textbf{11}, 53–57 (2017).



\bibitem{Averchenko2017}V. Averchenko, D. Sych, G. Schunk, U. Vogl, C. Marquardt, and G. Leuchs, \ShowTitle{Temporal shaping of single photons enabled by entanglement, }
Phys. Rev. A \textbf{96}, 043822 (2017).

\bibitem{Sych2017}D. Sych, V. Averchenko, and G. Leuchs, \ShowTitle{Generic method for lossless generation of arbitrarily shaped photons, }Phys. Rev. A \textbf{96}, 053847 (2017).

\bibitem{Reiserer2015} A. Reiserer and G. Rempe, \ShowTitle{Cavity-based quantum networks with single atoms and optical photons, }Rev. Mod. Phys. \textbf{87}, 1379 (2015) .

\bibitem{Law1997} C. K. Law and H. J. Kimble, \ShowTitle{Deterministic generation of a bit-stream of single-photon pulses, }J. Mod. Opt. \textbf{44}, 2067–2074 (1997).

\bibitem{Kuhn1999} A. Kuhn, M. Hennrich, T. Bondo, and G. Rempe, \ShowTitle{Controlled generation of single photons from a strongly coupled atom-cavity system, }Applied Physics B \textbf{69} (5-6), 373-377.

\bibitem{Muecke2013}M. M{\"u}cke, J. Bochmann, C. Hahn, A. Neuzner, C. N{\"o}lleke, A. Reiserer, G. Rempe, and S. Ritter, \ShowTitle{Generation of single photons from an atom-cavity system, }Phys. Rev. A \textbf{87}, 063805 (2013).

\bibitem{Fleischhauer2000} M. Fleischhauer, S. Yelin, and M. Lukin, \ShowTitle{How to trap photons? Storing single-photon quantum states in collective atomic excitations, }Opt. Commun. \textbf{179},
395 (2000).

\bibitem{Gorshkov2007}A. Gorshkov, A. Andr\'e, M. Lukin and A. S\o{}rensen, \ShowTitle{Photon storage in $\ensuremath{\Lambda}$-type optically dense atomic media. I. Cavity model, }Phys. Rev. A \textbf{76}, 033804(2007).

\bibitem{Dilley2012} J. Dilley, P. Nisbet-Jones, B. W. Shore, and A. Kuhn, \ShowTitle{Single-photon absorption in coupled atom-cavity systems, 
}Phys. Rev. A \textbf{85}, 023834 (2012).

\bibitem{Dalibard92}J. Dalibard, Y. Castin, and K. Mølmer, Wave-function approach to dissipative processes in quantum optics, Phys. Rev. Lett. \textbf{68}, 580 (1992)

\bibitem{Giannelli2018}L. Giannelli, T. Schmit, T. Calarco, C. P. Koch, S. Ritter, and G. Morigi, \ShowTitle{Optimal storage of a single photon by a single intra-cavity atom, }New J. Phys. \textbf{20}, 105009 (2018).



\bibitem{Morin2019}O. Morin, S. Langenfeld, M. Körber, and G. Rempe, \ShowTitle{Accurate photonic temporal mode analysis with reduced resources, } ArXiv 1909.00859.


\bibitem{Koerber2018} M. Körber, O. Morin, S. Langenfeld, A. Neuzner, S. Ritter, and G. Rempe,  \ShowTitle{Decoherence-protected memory for a single-photon qubit, }Nature Photon. \textbf{12}, 18-21 (2018).


\bibitem{CQEDNote}Notations of the parameters can vary from one reference to another. Here, $2\kappa$ is the full-width-half-maximum of the cavity spectrum (transmission or reflection). $1/2\gamma$ corresponds to the $1/e$ time of the excited state population decaying via spontaneous emission. $2g$ corresponds to the normal mode splitting that is observed when one atom is resonantly coupled to a cavity. $2\pi/\Omega$ is the period of population driving via Raman for instance. We provide again those definition in the SM.


\bibitem{MorinMode} O. Morin, C. Fabre, and J. Laurat, \ShowTitle{Experimentally accessing the optimal temporal mode of traveling quantum light states, }Phys. Rev. Lett. \textbf{111}, 213602 (2013).


\bibitem{LvovskyMode} Z. Qin, A. S. Prasad, T. Brannan, A. MacRae, A. Lezama, and A. I. Lvovsky, \ShowTitle{Complete temporal characterization of a single photon. }Light: Science \& Applications, \textbf{4}, e298 (2015).


\bibitem{Fid_temp_mode} The fidelity between two temporal modes is specified by $\mathcal{F}(e_1,e_2)=\abs{\int dt\  e_1(t) (e_2(t))^*}^2$. In our case, we compared the reconstructed temporal mode which can have complex values with the target one which only has real values.

\bibitem{Kurpiers2018} P. Kurpiers, P. Magnard, T. Walter, B. Royer, M. Pechal, J. Heinsoo, Y. Salathé, A. Akin, S. Storz, J.-C. Besse, S. Gasparinetti, A. Blais, and A. Wallraff, \ShowTitle{Deterministic quantum state transfer and remote entanglement using microwave photons, }Nature \textbf{558}, 264–267 (2018).


\bibitem{Enk97}S. J. van Enk, J. I. Cirac, and P. Zoller, Ideal Quantum Communication over Noisy Channels: A Quantum Optical Implementation, Phys. Rev. Lett. \textbf{78}, 4293 (1997).




\end{thebibliography}

\begin{thebibliography}{10}





\bibitem{Gorshkov}A. Gorshkov, A. Andr\'e, M. Lukin, and A. S\o{}rensen, Photon storage in $\ensuremath{\Lambda}$-type optically dense atomic media. I. Cavity model, Phys. Rev. A \textbf{76}, 033804(2007)

\bibitem{Dilley} J. Dilley, P. Nisbet-Jones, B. W. Shore, and A. Kuhn, Single-photon absorption in coupled atom-cavity systems, 
Phys. Rev. A \textbf{85}, 023834 (2012)

\bibitem{Luigi}L. Giannelli, T. Schmit, T. Calarco, C. P. Koch, S. Ritter, and G. Morigi, Optimal storage of a single photon by a single intra-cavity atom, New J. Phys \textbf{20}, 105009 (2018)

\bibitem{Dalibard}J. Dalibard, Y. Castin, and K. Mølmer, Wave-function approach to dissipative processes in quantum optics, Phys. Rev. Lett. \textbf{68}, 580 (1992)

\bibitem{Morin}O. Morin, S. Langenfeld, M. Körber, and G. Rempe, Accurate photonic temporal mode analysis with reduced resources, ArXiv 1909.00859


\end{thebibliography}

\end{document}